\documentclass[letterpaper,english,prl, showpacs,twocolumn]{revtex4}
\usepackage{mathptmx}
\usepackage[T1]{fontenc}
\usepackage[latin9]{inputenc}
\usepackage{soul}
\usepackage{xcolor}
\usepackage{graphicx}
\usepackage{amssymb}
\usepackage{esint}

\makeatletter

\newcommand{\noun}[1]{\textsc{#1}}

\providecolor{lyxadded}{rgb}{0,0,1}
\providecolor{lyxdeleted}{rgb}{1,0,0}

\usepackage{babel}
\makeatother

\begin{document}

\title{Spin Lifetimes in Quantum Dots from Noise Measurements}

\author{J. Wabnig}

\affiliation{Department of Materials, Oxford University, Oxford OX1 3PH, United
Kingdom}

\author{B. W. Lovett}

\affiliation{Department of Materials, Oxford University, Oxford OX1 3PH, United
Kingdom}

\author{J. H. Jefferson}

\affiliation{QinetiQ, St. Andrews Road, Malvern WR14 3PS, United Kingdom}

\author{G. A. D. Briggs}

\affiliation{Department of Materials, Oxford University, Oxford OX1 3PH, United
Kingdom}

\begin{abstract}
We present a method of obtaining information about spin lifetimes
in quantum dots from measurements of electrical transport. The dot
is under resonant microwave irradiation and at temperatures comparable
to or larger than the Zeeman energy. We find that the ratio of the
spin coherence times $T_{1}/T_{2}$ can be deduced from a measurement
of current through the quantum dot as a function the applied magnetic
field. We calculate the noise power spectrum of the dot current and
show that a dip occurs at the Rabi frequency with a line width given
by $1/T_{1}+1/T_{2}$.
\end{abstract}

\pacs{85.35.-p, 73.63.-b, 72.25.-b, 72.70.+m}

\maketitle
Electron spins are promising candidates for qubits \citep{LosDiV98}.
They constitute a natural two level system but in contrast with nuclear
spins they can be highly polarized simply by using routinely available
temperature and magnetic field conditions. Additionally one can exploit
the mobility of electrons, allowing spin manipulation and measurement
via charge currents, e.g. through tunnel barriers. 

For applications in quantum information processing it is essential
that spins retain phase information for as long as possible. This
is usually quantified by the spin coherence time $T_{2}$. Single
spin $T_{2}$ times have been determined optically \citep{hanson:161203,jelezko:076401}
and electrically \citep{koppens:766,Petta09302005}. Petta et al.
\citep{Petta09302005} encoded a qubit in the singlet and triplet
states of two electron on a double quantum dot and demonstrated coherent
manipulation, while Koppens et al. \citep{koppens:766} used pulsed
microwaves and top gates to demonstrate coherent oscillations of a
single spin in a double dot.

A method for determining spin coherence times using electrical transport
through a quantum dot in the stationary state has been suggested by
Engel and Loss \citep{engel:195321}, but this method can only determine
coherence times up to an upper limit that is related to the temperature
of the contacts. The study of current-current correlations has emerged
as a tool to detect coherence properties of quantum systems such as
quantum dots \citep{zhang:036603}. In nanomechanical resonators the
noise power spectrum has been used to determine the oscillator occupation
number and quality factor \citep{LaHaye04022004,flowers-jacobs:096804,brown:137205}.
The noise power spectrum of a current interacting with a charge qubit
\citep{barrett:017405,aguado:206601,korotkov:165310,ShnMozMar04}
and the current through a quantum dot under microwave radiation have
been thoroughly analyzed \citep{martin:018301,zhang:196602,dong:066601}.

In this Letter we suggest a method for measuring spin lifetimes using
current-current correlations. At experimentally accessible combinations
of temperature and magnetic field, where the Zeeman splitting is comparable
with or larger than the thermal energy, the measurement of arbitrary
intrinsic spin lifetimes becomes possible. To be able to detect the
influence of the finite electron spin lifetime on electronic transport
through the dot, the dwell time of an electron on the dot has to be
comparable to or longer than the spin relaxation times we wish to
detect, e.g. through a suitable choice of tunnel barriers. We will
henceforth assume this to be the case.

Our model consists of a quantum dot coupled weakly to two leads. The
full Hamiltonian after a rotating wave approximation reads $H=H_{D}+H_{E}+H_{T},$
where $H_{D}$ is the Hamiltonian of the dot, $H_{E}$ the Hamiltonian
for the leads and $H_{T}$ describes the tunneling between the leads
and the dot. %
\begin{figure}
\begin{centering}
\includegraphics[width=1\columnwidth]{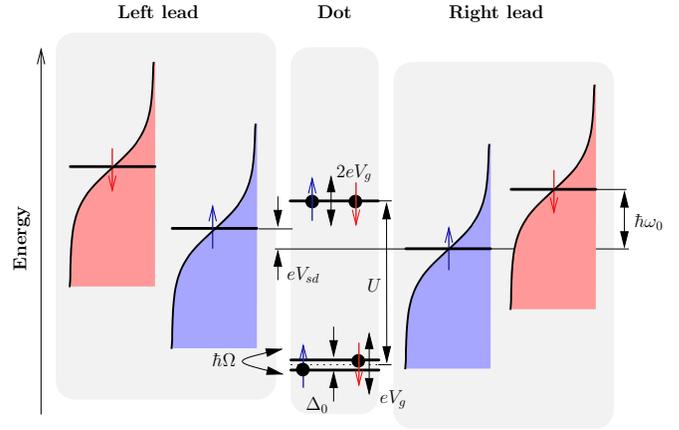}
\par\end{centering}

\caption{\label{fig1}(Color online) The level configuration on the quantum
dot and the distribution functions in the leads. The Fermi level of
the electrons in the leads is the same for spin up and spin down,
but in the rotating frame the energy of the electrons depends on their
spin orientation, and the microwave frequency $\omega_{0}$ determines
the difference in energy. Also the energy difference of the single
electron states on the dot is reduced in the rotating frame from the
Zeeman splitting $\Delta_{z}$ to the detuning $\Delta_{0}$, while
the energy of the two electron state is separated from the single
electron states by the charging energy $U$. The bias voltage $V_{sd}$
is applied symmetrically. The gate voltage $V_{g}$ shifts the energies
of the single electron levels by $eV_{g}$ and the two electron level
by $2eV_{g}$. The resonant Rabi frequency $\Omega$ couples spin
up and spin down.}

\end{figure}
We choose to model the quantum dot by including three many body states
on the quantum dot: the single electron states for spin up and spin
down respectively and a singlet two electron state. We assume that
the triplet two-electron states are separated from the singlet state
by a gap that is large compared to temperature and bias voltage. We
also do not take any additional single electron levels into account.
In a magnetic field $B$ only the single electron levels are split
by the Zeeman energy $\Delta_{z}=g\mu_{B}B$; the singlet is unaffected.
Under microwave irradiation with frequency $\omega_{0}$ and after
a rotating wave approximation the Hamiltonian for the dot can thus
be written \begin{eqnarray*}
H_{D} & = & \left(eV_{g}-\frac{\Delta_{0}}{2}\right)n_{\uparrow}+\left(eV_{g}+\frac{\Delta_{0}}{2}\right)n_{\downarrow}\\
 &  & +Un_{\uparrow}n_{\downarrow}+\frac{\hbar\Omega}{2}\left(d_{\downarrow}^{\dagger}d_{\uparrow}+d_{\uparrow}^{\dagger}d_{\downarrow}\right),\end{eqnarray*}
where $d_{s}^{\dagger}$ and $d_{s}$, $s=\uparrow,\downarrow$, are
the creation and annihilation operator for dot electrons in spin up
and spin down state respectively and $n_{s}=d_{s}^{\dagger}d_{s}$.
$U$ is the mutual Coulomb repulsion energy between two electrons
on the dot, $\Omega$ the resonant Rabi frequency, arising from an
oscillating magnetic field in x-direction, e.g. in the electric field
node of a microwave cavity, and $\Delta_{0}=\Delta_{z}-\hbar\omega_{0}$
is the detuning of the applied microwave field . We have also included
the action of a back gate that shifts the single electron energies
by the amount $eV_{g}$ (for the energy level diagram see Fig.~\ref{fig1}).
The Hamiltonian for the leads is given by\[
H_{E}=\sum_{k\alpha s}\varepsilon_{k\alpha s}c_{k\alpha s}^{\dagger}c_{k\alpha s},\]
where $c_{kas}^{\dagger},c_{kas}$, with $\alpha=L,R$ and $s=\uparrow,\downarrow$,
are the operators for the lead electrons with momentum $k$. Tunneling
from the leads to the dots is given by\[
H_{T}(t)=\sum_{k\alpha s}t_{\alpha}c_{k\alpha s}^{\dagger}(t)d_{s}+h.c.,\]
where $t_{a}$ are the tunneling amplitudes, which in the following
we assume to be independent of the electron energy and equal for both
leads. We have included the time dependence resulting from the transformation
to the rotating frame in the lead operators, introducing $c_{k\alpha\uparrow}^{\dagger}(t)=c_{k\alpha\uparrow}^{\dagger}e^{-i\omega_{0}t}$
and $c_{k\alpha\downarrow}^{\dagger}(t)=c_{k\alpha\downarrow}^{\dagger}e^{i\omega_{0}t}$.

To describe electronic transport through the dot we adapt a Markovian
master equation approach \citep{wabnig:165347}. The reduced density
matrix of the dot is defined as $\rho(t)=\textnormal{Tr}_{E}\,\left[R(t)\right]$,
where $\textnormal{Tr}_{E}\left[\ldots\right]$ denotes the trace
over the environment and $R$ is the the density matrix of the full
system. Its evolution is given by the master equation %
\begin{figure}
\begin{centering}
\includegraphics[width=0.8\columnwidth]{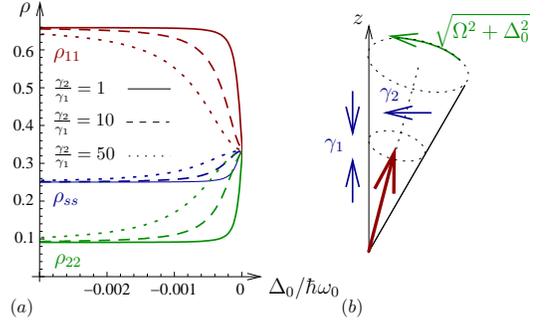}
\par\end{centering}

\caption{\label{fig2}(Color online) (a) The diagonal elements of the density
matrix in the energy basis as a function of the detuning for different
ratios of the decay rates $\gamma_{2}/\gamma_{1}$ and $\gamma_{e}=10^{-7}\omega_{0}$,
$\gamma_{1}=10^{-6}\omega_{0}$, $\Omega=10^{-4}\omega_{0}$, $k_{B}T=\hbar\omega_{0}$,
$eV_{sd}=\hbar\omega_{0}$ and $V_{g}=0$. The dotted lines correspond
to $\gamma_{2}/\gamma_{1}=50$, the dashed to $\gamma_{2}/\gamma_{1}=10$
and the solid lines to $\gamma_{2}/\gamma_{1}=1$. (b) Diagram illustrating
the influence of the different decoherence mechanisms on the spin
dynamics.}

\end{figure}
$d\rho(t)/dt=\mathcal{K}\cdot\rho(t),$ where the super-operator $\mathcal{K}$
consists of three parts: $\mathcal{K}=\mathcal{K}_{0}+\mathcal{K}_{e}+\mathcal{K}_{int}.$
The part describing the free evolution is given by\[
\mathcal{K}_{0}\cdot X=-i\hbar\left[H_{D},X\right],\]
where $X$ can be any operator. Renormalization effects due to interaction
with the environment are included in the free evolution (see \citep{wabnig:165347}).
We can also define the free propagator in Fourier space\[
\mathcal{U}_{0}(\varepsilon)=\int dt\, e^{\mathcal{K}_{0}t}\, e^{i\varepsilon t/\hbar}.\]
The influence of the electrons tunneling from the leads is described
by the super-operator\begin{eqnarray*}
\mathcal{K}_{e} & = & \gamma_{e}\int d\varepsilon\sum_{\alpha sij}s_{ij}\left[g^{ij}(\varepsilon-\mu_{\alpha s})\mathcal{A}_{s}^{(i)\dagger}\cdot\mathcal{U}_{0}(\varepsilon)\cdot\mathcal{A}_{s}^{(j)}\right.\\
 &  & +\left.g^{ij}(\varepsilon+\mu_{\alpha s})\mathcal{A}_{s}^{(i)}\cdot\mathcal{U}_{0}(\varepsilon)\cdot\mathcal{A}_{s}^{(j)\dagger}\right]\end{eqnarray*}
where $\gamma_{e}=2\pi g(E_{F})\left|t_{L}\right|^{2}$, $g(E_{F})$
being the density of states at the Fermi energy, $s_{ij}=1$ for $i=j$
and $s_{ij}=-1$ otherwise. $\mu_{L/R,\uparrow/\downarrow}=\mu_{L/R}\mp\hbar\omega_{0}/2$,
$\mu_{L/R}=\mu\pm eV_{sd}/2$, $g^{11}(\varepsilon)=g^{12}(\varepsilon)=1-f(\varepsilon)$,
$g^{21}(\varepsilon)=g^{22}(\varepsilon)=f(\varepsilon)$, $f(\varepsilon)$being
the Fermi function, $\mathcal{A}_{s}^{(1)}\cdot X=d_{s}X$ and $\mathcal{A}_{s}^{(2)}\cdot X=Xd_{s}$,
and similarly for the hermitian conjugate. The use of the free propagator
instead of the self consistent inclusion of the full propagator does
not alter the result as long as the distribution functions $g^{ij}$
are smooth on the scale where the full propagator is peaked. In addition
to tunneling, we include a phenomenological description of the intrinsic
spin decay, given by \[
\mathcal{K}_{int}=\mathcal{K}_{int}^{(1)}+\mathcal{K}_{int}^{(2)}.\]
The first part describes an energy relaxation process leading towards
thermal equilibrium $\mathcal{K}_{int}^{(1)}\cdot X=\gamma_{+}\left(\sigma_{+}\sigma_{-}X+X\sigma_{+}\sigma_{-}-\sigma_{+}X\sigma_{-}\right)+\gamma_{-}\left(\sigma_{-}\sigma_{+}X+X\sigma_{-}\sigma_{+}-\sigma_{-}X\sigma_{+}\right)$
with $\gamma_{+}+\gamma_{-}=\gamma_{1}=1/T_{1}$ and $\gamma_{+}/\gamma_{-}=\exp(\Delta_{z}/k_{B}T)$
where $\sigma_{+}=d_{\downarrow}^{\dagger}d_{\uparrow}$, $\sigma_{-}=d_{\uparrow}^{\dagger}d_{\downarrow}$
and $\sigma_{z}=d_{\downarrow}^{\dagger}d_{\downarrow}-d_{\uparrow}^{\dagger}d_{\uparrow}$.
The second part is a pure dephasing process $\mathcal{K}_{int}^{(2)}\cdot X=\left(\gamma_{2}-\gamma_{1}/2\right)\left(X-\sigma_{z}X\sigma_{z}\right),$
where $\gamma_{2}=1/T_{2}$. We will be interested in the electronic
transport through the dot in the stationary state. The stationary
state density matrix can be derived from $\mathcal{K}\cdot\rho_{st}=0.$
This equation was solved analytically using \noun{Mathematica}. For
small detunings $|\Delta_{0}|\ll\hbar\omega_{0}$ and $\hbar\omega_{0}<k_{B}T$
the density matrix elements in the energy basis have the following
approximate dependence on the detuning\begin{equation}
\rho_{ii}(\Delta_{0})=\left\langle i\right|\rho_{st}\left|i\right\rangle =\frac{\rho_{ii}^{(\infty)}\Delta_{0}^{2}+\rho_{ii}^{(0)}\Omega^{2}\gamma_{2}/\gamma_{1}}{\Delta_{0}^{2}+\Omega^{2}\gamma_{2}/\gamma_{1}},\label{eq:population}\end{equation}
where $\rho_{ii}^{(\infty)}$ denotes the density matrix element far
away from resonance and $\rho_{ii}^{(0)}$ the density matrix element
on resonance. The dependence of the density matrix elements on the
magnetic field for $V_{g}=0$ is shown in Fig.~\ref{fig2}(a). Far
off resonance the ratio between the spin up and spin down populations
is thermal while on resonance the three populations equalize. The
width of the transition from the thermal state to the equalized state
is given by $\Omega\sqrt{\gamma_{2}/\gamma_{1}}$. This can be understood
with the help of Fig.~\ref{fig2}(b): The spin precesses around the
direction of the effective magnetic field, $\mathbf{B}/g\mu_{B}=(\Omega,0,\Delta_{0})$.
For the parts of the precession closer to the z-axis the $\gamma_{1}$
processes drive the spin towards a thermal state along the z-axis
of the Bloch sphere. Further away from the z-axis the $\gamma_{2}$
processes dominate, tending to reduce the x and y component to zero.
The precession mixes both processes and the relative rates of both
relaxation processes determines the stationary state. For $\gamma_{2}>\gamma_{1}$
the spin reacts to a different direction in the magnetic field further
from resonance, resulting in a wider transition from thermal to equalized
population. 

\begin{figure}
\begin{centering}
\includegraphics[width=1\columnwidth]{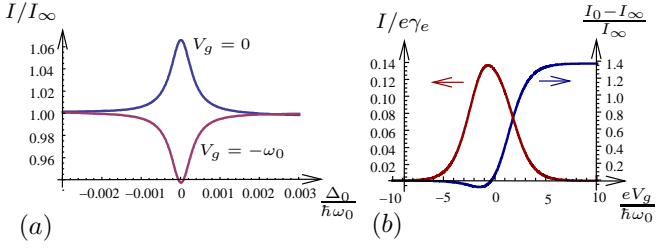}
\par\end{centering}

\caption{\label{fig3}(Color online) (a) The current $I$ as a function of
the detuning relative to the current at $\Delta_{0}=0$ without microwave
irradiation for different gate voltages and $eV_{sd}=\hbar\omega_{0}$,
$k_{B}T=\hbar\omega_{0}$, $\Omega=10^{-4}\omega_{0}$, $\gamma_{e}=10^{-7}\omega_{0}$,
$\gamma_{1}=\gamma_{2}=10^{-6}\omega_{0}$. When the magnetic field
matches the microwave frequency the current shows a peak or a dip
depending on the applied gate voltage. (b) The current at resonance,
$I_{0}$, relative to the current without microwave irradiation and
the current as a function of the gate voltage, otherwise same values
as in (a). }

\end{figure}
The current through the dot in the stationary state can be expressed
as $I=e\textnormal{Tr}_{D}\left[\mathcal{J}\cdot\rho_{st}\right],$
where $\textnormal{Tr}_{D}\left[\ldots\right]$ denotes the trace
over the dot degrees of freedom, with the current super-operator given
by\begin{eqnarray*}
\mathcal{J} & = & \gamma_{e}\int d\varepsilon\sum_{\alpha s}\left[g^{ij}(\varepsilon-\mu_{\alpha s})\mathcal{A}_{\alpha s}^{(j)\dagger}\cdot\mathcal{U}_{0}(\varepsilon)\cdot\mathcal{A}_{\alpha s}^{(i)}\right.\\
 &  & -\left.g^{ji}(\varepsilon+\mu_{\alpha s})\mathcal{A}_{\alpha s}^{(j)}\cdot\mathcal{U}_{0}(\varepsilon)\cdot\mathcal{A}_{\alpha s}^{(i)\dagger}\right],\, i\neq j,\, i,j=1,2.\end{eqnarray*}
For equal tunneling rates for both leads it is sufficient to consider
only the current through one of the tunneling contacts (L or R). In
the region of small detunings, $|\Delta_{0}|\ll V_{sd},\, T$, the
current can be written approximately as $I=i_{11}\rho_{11}+i_{22}\rho_{22}+i_{33}\rho_{33},$
where the currents $i_{kk},\, k=1,2,3$ depend on $V_{sd},\, V_{g},\, T$
but only weakly on the detuning. The magnetic field dependence of
the stationary current for different gate voltages is shown in Fig.~\ref{fig3}.
Close to the resonance the current shows a peak or a dip, depending
on the bias voltage. The dependence of the current on the detuning
follows from the detuning dependence of the density matrix elements
as\[
I=\frac{I_{\infty}\Delta_{0}^{2}+I_{0}\Omega^{2}\gamma_{2}/\gamma_{1}}{\Delta_{0}^{2}+\Omega^{2}\gamma_{2}/\gamma_{1}},\]
where $I_{\infty}$ is the current far away from the resonance and
$I_{0}$ the current at resonance. It is therefore possible to extract
the ratio of the dephasing rates $\gamma_{1}/\gamma_{2}$ from the
width of the current peak if the Rabi frequency $\Omega$ is known.

The noise power spectrum consists of a constant noise floor, $S_{F}$,
and two features, $S_{S}(\omega)$, sitting on top of that noise floor,
a peak at zero frequency and a%
\begin{figure}
\begin{centering}
\includegraphics[width=1\columnwidth]{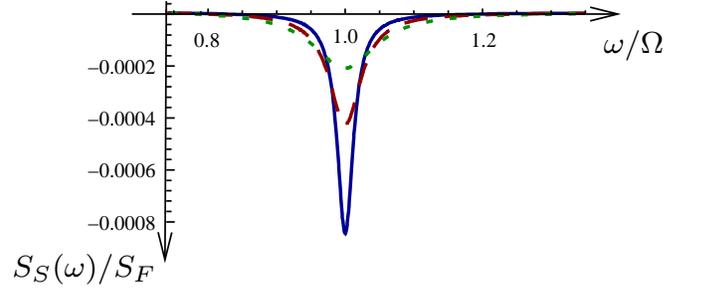}
\par\end{centering}

\caption{\label{fig4}(Color online) The peaked part of noise power spectrum,
$S_{S}(\omega)$, relative to the noise floor around the Rabi frequency
at resonance ($\Delta_{z}=\hbar\omega_{0})$ and for different rates
$\gamma_{1}$ and $V_{g}=0$ , $eV_{sd}=\hbar\omega_{0}$, $k_{B}T=\hbar\omega_{0}$,
$\Omega=10^{-4}\omega_{0}$, $\gamma_{e}=10^{-7}\omega_{0}$ and a
fixed ratio $\gamma_{2}/\gamma_{1}=10$. The solid line corresponds
to $\gamma_{1}=5\cdot10^{-7}\omega_{0}$, the dashed line to $\gamma_{1}=10^{-6}\omega_{0}$
and the dotted line to $\gamma_{1}=2\cdot10^{-6}\omega_{0}$. }

\end{figure}
 dip at the Rabi frequency. A similar noise power spectrum was discussed
by Aguado and Brandes for a charge qubit in \citep{aguado:206601}.
The floor is given by \[
S_{F}=e^{2}\textnormal{Tr}_{D}\left[\mathcal{D}\cdot\rho_{st}\right],\]
where we introduced the diffusion super-operator \begin{eqnarray*}
\mathcal{D} & = & \gamma_{e}\int d\varepsilon\sum_{\alpha s}\left[g^{ij}(\varepsilon-\mu_{\alpha s})\mathcal{A}_{\alpha s}^{(j)\dagger}\cdot\mathcal{U}_{0}(\varepsilon)\cdot\mathcal{A}_{\alpha s}^{(i)}\right.\\
 &  & +\left.g^{ji}(\varepsilon+\mu_{\alpha s})\mathcal{A}_{\alpha s}^{(j)}\cdot\mathcal{U}_{0}(\varepsilon)\cdot\mathcal{A}_{\alpha s}^{(i)\dagger}\right],\, i\neq j,\, i,j=1,2.\end{eqnarray*}
and the features in the noise power spectrum are described by the
term\[
S_{S}(\omega)=\textnormal{Tr}_{D}\left[\delta\mathcal{J}\cdot\mathcal{U}(\omega)\cdot\delta\mathcal{J}\cdot\rho_{st}\right],\]
with $\delta\mathcal{J}=\mathcal{J}-I$ and the evolution super operator\[
\mathcal{U}(\varepsilon)=\int dt\, e^{\mathcal{K}t}\, e^{i\varepsilon t/\hbar}.\]
This frequency dependence of the features in the noise power spectrum
can be evaluated at resonance to give in the positive half of the
spectrum \[
S_{S}(\omega)=S_{0}\frac{\gamma_{e}^{2}}{\omega^{2}+\gamma_{e}^{2}}-S_{\Omega}\frac{\gamma_{\Omega}^{2}}{\left(\omega-\Omega\right)^{2}+\gamma_{\Omega}^{2}},\]
where $S_{0/\Omega}$ are the peak heights at zero and at the Rabi
frequency. The peak and the dip have both a Lorentzian line-shape.
The peak at zero frequency stems from elastic processes on the dot,
its width is determined solely by the tunneling rate $\gamma_{e}$,
while the dip appearing at the Rabi frequency is due to processes
involving energy exchange on the dot and its width depends on the
average of the two spin decoherence rates, $\gamma_{\Omega}=(\gamma_{1}+\gamma_{2})/2$.

We have now all the ingredients to deduce the spin lifetimes $T_{1}$
and $T_{2}$. We can envisage the following measurement procedure:
A scan of the gate voltage without an applied microwave field will
produce a series of resonances in the current. Selecting one of the
resonances and scanning the magnetic field while applying microwave
radiation will give information about the presence of an active spin
on the dot and the width of the resulting peak gives the ratio $T_{1}/T_{2}$.
Finally adjusting the magnetic field to be exactly on resonance and
measuring the noise power spectrum around the Rabi frequency we can
obtain $1/T_{1}+1/T_{2}$, thus making it possible to deduce $T_{1}$
and $T_{2}$. 

Why is is necessary to measure the current noise spectrum at finite
frequency? Engel and Loss showed that measuring only the current through
a quantum dot as a function of the gate voltage can already give information
about $T_{2}$, but to resolve a certain line-width in the dot the
distribution functions of the leads have to be sharper than the line-width
of the level under consideration. The typical sharpness of the distribution
function is given by temperature. The achievable temperatures in today's
dilution refrigerators of $\sim10\,\textnormal{mK}$ correspond to
a resolvable line-width of $\sim100\,\textnormal{MHz}$, while values
for $T_{1/2}$ in group IV semiconductors correspond to line-widths
of $<100\,\textnormal{Hz}$ \citep{TyrMorBen06}. Our scheme will
work even at temperatures high compared to the intrinsic line-widths
of the dot levels and works well at temperatures comparable to or
larger than the Zeeman splitting of the single electron levels, although
for temperatures much larger than the Zeeman splitting signal to noise
ratio deteriorates. The method operates in the steady state and no
pulsed gates are necessary.

What are the limitations of our method? To detect a certain intrinsic
decay rate the tunneling rate has to be smaller than that rate. For
expected decay rates of the order of $100\,\textnormal{Hz}$ that
would correspond to a current of $0.01\,\textnormal{fA}$. Currents
of this magnitude are measurable using electron counting techniques
\citep{lu:422}. At the same time we need to detect the noise power
spectrum around the Rabi frequency, typically of the order of MHz.
To resolve the peak the charge counting has thus to work at rates
of $10\,\textnormal{MHz}$ or higher. Radio frequency (RF) single
electron transistors or RF quantum point contacts have been shown
to work as charge counters up to 1 GHz and should make such measurements
possible.

In our modeling we phenomenologically introduced the spin decay rates
as constants. A more detailed analysis of the decoherence mechanisms
shows that the coherence time $T_{2}$ will, in principle, depend
on the Rabi frequency. Assuming a Markovian bath the decay rate is
directly related to the spectral density of the bath at the Rabi frequency.
Changing the Rabi frequency by altering the microwave intensity should
make it therefore possible to probe the spectral density of the bath.
A dependence of the coherence time on microwave intensity has been
measured in \citep{koppens:766}. With this simple model we hope to
have shown the usefulness of noise measurements in determining spin
lifetimes.

\begin{acknowledgments}
This work is part of QIP IRC. JW thanks The Wenner-Gren Foundations
for financial support. BWL acknowledges support from the Royal Society.
JHJ acknowledges support from the UK MOD. GADB is supported by an
EPSRC Professional Research Fellowship. \bibliographystyle{apsrev}
\bibliography{esrnoiseNEW3}

\begin{thebibliography}{20}
\expandafter\ifx\csname natexlab\endcsname\relax\def\natexlab#1{#1}\fi
\expandafter\ifx\csname bibnamefont\endcsname\relax
  \def\bibnamefont#1{#1}\fi
\expandafter\ifx\csname bibfnamefont\endcsname\relax
  \def\bibfnamefont#1{#1}\fi
\expandafter\ifx\csname citenamefont\endcsname\relax
  \def\citenamefont#1{#1}\fi
\expandafter\ifx\csname url\endcsname\relax
  \def\url#1{\texttt{#1}}\fi
\expandafter\ifx\csname urlprefix\endcsname\relax\def\urlprefix{URL }\fi
\providecommand{\bibinfo}[2]{#2}
\providecommand{\eprint}[2][]{\url{#2}}

\bibitem[{\citenamefont{Loss and DiVincenzo}(1998)}]{LosDiV98}
\bibinfo{author}{\bibfnamefont{D.}~\bibnamefont{Loss}} \bibnamefont{and}
  \bibinfo{author}{\bibfnamefont{D.~P.} \bibnamefont{DiVincenzo}},
  \bibinfo{journal}{Phys. Rev. A} \textbf{\bibinfo{volume}{57}},
  \bibinfo{pages}{120} (\bibinfo{year}{1998}).

\bibitem[{\citenamefont{Hanson et~al.}(2006)\citenamefont{Hanson, Gywat, and
  Awschalom}}]{hanson:161203}
\bibinfo{author}{\bibfnamefont{R.}~\bibnamefont{Hanson}},
  \bibinfo{author}{\bibfnamefont{O.}~\bibnamefont{Gywat}}, \bibnamefont{and}
  \bibinfo{author}{\bibfnamefont{D.~D.} \bibnamefont{Awschalom}},
  \bibinfo{journal}{Phys. Rev. B} \textbf{\bibinfo{volume}{74}},
  \bibinfo{pages}{161203(R)} (\bibinfo{year}{2006}).

\bibitem[{\citenamefont{Jelezko et~al.}(2004)\citenamefont{Jelezko, Gaebel,
  Popa, Gruber, and Wrachtrup}}]{jelezko:076401}
\bibinfo{author}{\bibfnamefont{F.}~\bibnamefont{Jelezko}},
  \bibinfo{author}{\bibfnamefont{T.}~\bibnamefont{Gaebel}},
  \bibinfo{author}{\bibfnamefont{I.}~\bibnamefont{Popa}},
  \bibinfo{author}{\bibfnamefont{A.}~\bibnamefont{Gruber}}, \bibnamefont{and}
  \bibinfo{author}{\bibfnamefont{J.}~\bibnamefont{Wrachtrup}},
  \bibinfo{journal}{Phys. Rev. Lett.} \textbf{\bibinfo{volume}{92}},
  \bibinfo{pages}{076401} (\bibinfo{year}{2004}).

\bibitem[{\citenamefont{Koppens et~al.}(2006)\citenamefont{Koppens, Buizert,
  Tielrooij, Vink, Nowack, Meunier, Kouwenhoven, and
  Vandersypen}}]{koppens:766}
\bibinfo{author}{\bibfnamefont{F.~H.~L.} \bibnamefont{Koppens}},
  \bibinfo{author}{\bibfnamefont{C.}~\bibnamefont{Buizert}},
  \bibinfo{author}{\bibfnamefont{K.~J.} \bibnamefont{Tielrooij}},
  \bibinfo{author}{\bibfnamefont{I.~T.} \bibnamefont{Vink}},
  \bibinfo{author}{\bibfnamefont{K.~C.} \bibnamefont{Nowack}},
  \bibinfo{author}{\bibfnamefont{T.}~\bibnamefont{Meunier}},
  \bibinfo{author}{\bibfnamefont{L.~P.} \bibnamefont{Kouwenhoven}},
  \bibnamefont{and} \bibinfo{author}{\bibfnamefont{L.~M.~K.}
  \bibnamefont{Vandersypen}}, \bibinfo{journal}{Nature}
  \textbf{\bibinfo{volume}{442}}, \bibinfo{pages}{766} (\bibinfo{year}{2006}).

\bibitem[{\citenamefont{Petta et~al.}(2005)\citenamefont{Petta, Johnson,
  Taylor, Laird, Yacoby, Lukin, Marcus, Hanson, and Gossard}}]{Petta09302005}
\bibinfo{author}{\bibfnamefont{J.~R.} \bibnamefont{Petta}},
  \bibinfo{author}{\bibfnamefont{A.~C.} \bibnamefont{Johnson}},
  \bibinfo{author}{\bibfnamefont{J.~M.} \bibnamefont{Taylor}},
  \bibinfo{author}{\bibfnamefont{E.~A.} \bibnamefont{Laird}},
  \bibinfo{author}{\bibfnamefont{A.}~\bibnamefont{Yacoby}},
  \bibinfo{author}{\bibfnamefont{M.~D.} \bibnamefont{Lukin}},
  \bibinfo{author}{\bibfnamefont{C.~M.} \bibnamefont{Marcus}},
  \bibinfo{author}{\bibfnamefont{M.~P.} \bibnamefont{Hanson}},
  \bibnamefont{and} \bibinfo{author}{\bibfnamefont{A.~C.}
  \bibnamefont{Gossard}}, \bibinfo{journal}{Science}
  \textbf{\bibinfo{volume}{309}}, \bibinfo{pages}{2180} (\bibinfo{year}{2005}).

\bibitem[{\citenamefont{Engel and Loss}(2002)}]{engel:195321}
\bibinfo{author}{\bibfnamefont{H.-A.} \bibnamefont{Engel}} \bibnamefont{and}
  \bibinfo{author}{\bibfnamefont{D.}~\bibnamefont{Loss}},
  \bibinfo{journal}{Phys. Rev. B} \textbf{\bibinfo{volume}{65}},
  \bibinfo{eid}{195321} (\bibinfo{year}{2002}).

\bibitem[{\citenamefont{Zhang et~al.}(2007)\citenamefont{Zhang, DiCarlo,
  McClure, Yamamoto, Tarucha, Marcus, Hanson, and Gossard}}]{zhang:036603}
\bibinfo{author}{\bibfnamefont{Y.}~\bibnamefont{Zhang}},
  \bibinfo{author}{\bibfnamefont{L.}~\bibnamefont{DiCarlo}},
  \bibinfo{author}{\bibfnamefont{D.~T.} \bibnamefont{McClure}},
  \bibinfo{author}{\bibfnamefont{M.}~\bibnamefont{Yamamoto}},
  \bibinfo{author}{\bibfnamefont{S.}~\bibnamefont{Tarucha}},
  \bibinfo{author}{\bibfnamefont{C.~M.} \bibnamefont{Marcus}},
  \bibinfo{author}{\bibfnamefont{M.~P.} \bibnamefont{Hanson}},
  \bibnamefont{and} \bibinfo{author}{\bibfnamefont{A.~C.}
  \bibnamefont{Gossard}}, \bibinfo{journal}{Phys. Rev. Lett.}
  \textbf{\bibinfo{volume}{99}}, \bibinfo{pages}{036603}
  (\bibinfo{year}{2007}).

\bibitem[{\citenamefont{LaHaye et~al.}(2004)\citenamefont{LaHaye, Buu,
  Camarota, and Schwab}}]{LaHaye04022004}
\bibinfo{author}{\bibfnamefont{M.~D.} \bibnamefont{LaHaye}},
  \bibinfo{author}{\bibfnamefont{O.}~\bibnamefont{Buu}},
  \bibinfo{author}{\bibfnamefont{B.}~\bibnamefont{Camarota}}, \bibnamefont{and}
  \bibinfo{author}{\bibfnamefont{K.~C.} \bibnamefont{Schwab}},
  \bibinfo{journal}{Science} \textbf{\bibinfo{volume}{304}},
  \bibinfo{pages}{74} (\bibinfo{year}{2004}).

\bibitem[{\citenamefont{Flowers-Jacobs
  et~al.}(2007)\citenamefont{Flowers-Jacobs, Schmidt, and
  Lehnert}}]{flowers-jacobs:096804}
\bibinfo{author}{\bibfnamefont{N.~E.} \bibnamefont{Flowers-Jacobs}},
  \bibinfo{author}{\bibfnamefont{D.~R.} \bibnamefont{Schmidt}},
  \bibnamefont{and} \bibinfo{author}{\bibfnamefont{K.~W.}
  \bibnamefont{Lehnert}}, \bibinfo{journal}{Phys. Rev. Lett.}
  \textbf{\bibinfo{volume}{98}}, \bibinfo{pages}{096804}
  (\bibinfo{year}{2007}).

\bibitem[{\citenamefont{Brown et~al.}(2007)\citenamefont{Brown, Britton,
  Epstein, Chiaverini, Leibfried, and Wineland}}]{brown:137205}
\bibinfo{author}{\bibfnamefont{K.~R.} \bibnamefont{Brown}},
  \bibinfo{author}{\bibfnamefont{J.}~\bibnamefont{Britton}},
  \bibinfo{author}{\bibfnamefont{R.~J.} \bibnamefont{Epstein}},
  \bibinfo{author}{\bibfnamefont{J.}~\bibnamefont{Chiaverini}},
  \bibinfo{author}{\bibfnamefont{D.}~\bibnamefont{Leibfried}},
  \bibnamefont{and} \bibinfo{author}{\bibfnamefont{D.~J.}
  \bibnamefont{Wineland}}, \bibinfo{journal}{Phys. Rev. Lett.}
  \textbf{\bibinfo{volume}{99}}, \bibinfo{eid}{137205} (\bibinfo{year}{2007}).

\bibitem[{\citenamefont{Barrett and Stace}(2006)}]{barrett:017405}
\bibinfo{author}{\bibfnamefont{S.~D.} \bibnamefont{Barrett}} \bibnamefont{and}
  \bibinfo{author}{\bibfnamefont{T.~M.} \bibnamefont{Stace}},
  \bibinfo{journal}{Phys. Rev. Lett.} \textbf{\bibinfo{volume}{96}},
  \bibinfo{eid}{017405} (\bibinfo{year}{2006}).

\bibitem[{\citenamefont{Aguado and Brandes}(2004)}]{aguado:206601}
\bibinfo{author}{\bibfnamefont{R.}~\bibnamefont{Aguado}} \bibnamefont{and}
  \bibinfo{author}{\bibfnamefont{T.}~\bibnamefont{Brandes}},
  \bibinfo{journal}{Phys. Rev. Lett.} \textbf{\bibinfo{volume}{92}},
  \bibinfo{pages}{206601} (\bibinfo{year}{2004}).

\bibitem[{\citenamefont{Korotkov and Averin}(2001)}]{korotkov:165310}
\bibinfo{author}{\bibfnamefont{A.~N.} \bibnamefont{Korotkov}} \bibnamefont{and}
  \bibinfo{author}{\bibfnamefont{D.~V.} \bibnamefont{Averin}},
  \bibinfo{journal}{Phys. Rev. B} \textbf{\bibinfo{volume}{64}},
  \bibinfo{pages}{165310} (\bibinfo{year}{2001}).

\bibitem[{\citenamefont{Shnirman et~al.}(2004)\citenamefont{Shnirman, Mozyrsky,
  and Martin}}]{ShnMozMar04}
\bibinfo{author}{\bibfnamefont{A.}~\bibnamefont{Shnirman}},
  \bibinfo{author}{\bibfnamefont{D.}~\bibnamefont{Mozyrsky}}, \bibnamefont{and}
  \bibinfo{author}{\bibfnamefont{I.}~\bibnamefont{Martin}},
  \bibinfo{journal}{EPL (Europhysics Letters)} \textbf{\bibinfo{volume}{67}},
  \bibinfo{pages}{840} (\bibinfo{year}{2004}).

\bibitem[{\citenamefont{Martin et~al.}(2003)\citenamefont{Martin, Mozyrsky, and
  Jiang}}]{martin:018301}
\bibinfo{author}{\bibfnamefont{I.}~\bibnamefont{Martin}},
  \bibinfo{author}{\bibfnamefont{D.}~\bibnamefont{Mozyrsky}}, \bibnamefont{and}
  \bibinfo{author}{\bibfnamefont{H.~W.} \bibnamefont{Jiang}},
  \bibinfo{journal}{Phys. Rev. Lett.} \textbf{\bibinfo{volume}{90}},
  \bibinfo{eid}{018301} (\bibinfo{year}{2003}).

\bibitem[{\citenamefont{Zhang et~al.}(2003)\citenamefont{Zhang, Xue, and
  Xie}}]{zhang:196602}
\bibinfo{author}{\bibfnamefont{P.}~\bibnamefont{Zhang}},
  \bibinfo{author}{\bibfnamefont{Q.-K.} \bibnamefont{Xue}}, \bibnamefont{and}
  \bibinfo{author}{\bibfnamefont{X.~C.} \bibnamefont{Xie}},
  \bibinfo{journal}{Phys. Rev. Lett.} \textbf{\bibinfo{volume}{91}},
  \bibinfo{pages}{196602} (\bibinfo{year}{2003}).

\bibitem[{\citenamefont{Dong et~al.}(2005)\citenamefont{Dong, Cui, and
  Lei}}]{dong:066601}
\bibinfo{author}{\bibfnamefont{B.}~\bibnamefont{Dong}},
  \bibinfo{author}{\bibfnamefont{H.~L.} \bibnamefont{Cui}}, \bibnamefont{and}
  \bibinfo{author}{\bibfnamefont{X.~L.} \bibnamefont{Lei}},
  \bibinfo{journal}{Phys. Rev. Lett.} \textbf{\bibinfo{volume}{94}},
  \bibinfo{pages}{066601} (\bibinfo{year}{2005}).

\bibitem[{\citenamefont{Wabnig et~al.}(2005)\citenamefont{Wabnig, Khomitsky,
  Rammer, and Shelankov}}]{wabnig:165347}
\bibinfo{author}{\bibfnamefont{J.}~\bibnamefont{Wabnig}},
  \bibinfo{author}{\bibfnamefont{D.~V.} \bibnamefont{Khomitsky}},
  \bibinfo{author}{\bibfnamefont{J.}~\bibnamefont{Rammer}}, \bibnamefont{and}
  \bibinfo{author}{\bibfnamefont{A.~L.} \bibnamefont{Shelankov}},
  \bibinfo{journal}{Phys. Rev. B} \textbf{\bibinfo{volume}{72}},
  \bibinfo{pages}{165347} (\bibinfo{year}{2005}).

\bibitem[{\citenamefont{Tyryshkin et~al.}(2006)\citenamefont{Tyryshkin, Morton,
  Benjamin, Ardavan, Briggs, Ager, and Lyon}}]{TyrMorBen06}
\bibinfo{author}{\bibfnamefont{A.~M.} \bibnamefont{Tyryshkin}},
  \bibinfo{author}{\bibfnamefont{J.~J.~L.} \bibnamefont{Morton}},
  \bibinfo{author}{\bibfnamefont{S.~C.} \bibnamefont{Benjamin}},
  \bibinfo{author}{\bibfnamefont{A.}~\bibnamefont{Ardavan}},
  \bibinfo{author}{\bibfnamefont{G.~A.~D.} \bibnamefont{Briggs}},
  \bibinfo{author}{\bibfnamefont{J.~W.} \bibnamefont{Ager}}, \bibnamefont{and}
  \bibinfo{author}{\bibfnamefont{S.~A.} \bibnamefont{Lyon}},
  \bibinfo{journal}{J. Phys.: Cond. Matt.} \textbf{\bibinfo{volume}{18}},
  \bibinfo{pages}{S783} (\bibinfo{year}{2006}).

\bibitem[{\citenamefont{Lu et~al.}(2003)\citenamefont{Lu, Ji, Pfeiffer, West,
  and Rimberg}}]{lu:422}
\bibinfo{author}{\bibfnamefont{W.}~\bibnamefont{Lu}},
  \bibinfo{author}{\bibfnamefont{Z.}~\bibnamefont{Ji}},
  \bibinfo{author}{\bibfnamefont{L.}~\bibnamefont{Pfeiffer}},
  \bibinfo{author}{\bibfnamefont{K.~W.} \bibnamefont{West}}, \bibnamefont{and}
  \bibinfo{author}{\bibfnamefont{A.~J.} \bibnamefont{Rimberg}},
  \bibinfo{journal}{Nature} \textbf{\bibinfo{volume}{423}},
  \bibinfo{pages}{422} (\bibinfo{year}{2003}).

\end{thebibliography}

\end{acknowledgments}

\end{document}